# A Multilevel Multinomial Logistic Regression Model for Identifying Risk Factors of Anemia in Children Aged 6-59 months in Northeastern States of India

## Sanku Dey[1] and Enayetur Raheem[2]

## Abstract

In this article, we use multilevel multinomial logistic regression model to identify the risk factors of anemia in children of northeastern States of India. The data consisted of 10,136 children of age group 6-59 months. We considered the level of anemia as the outcome variable with four ordinal categories (severe, moderate, mild, and non-anemic) based on hemoglobin concentration in blood as per WHO guidelines. A two-level random intercept model was considered with state of residence as the level-2 variable. The intra-class correlation (ICC) between states is 0.0577 indicating approximately 6% of the total variation in the response variable accounted for by the state of residence. Several multilevel models have been compared, and a final model was decided based on deviance test. We observed that predicted probability of being at or below severely anemic level to be 0.1247, at moderately anemic level: 0.3578, at mildly anemic level: 0.0698, and being non-anemic to be 0.4477. We found that age at marriage (OR=1.13, 95% CI: 1.05, 1.21) and the number of children even born (OR=1.09, 95% CI: 1.03, 1.15) have significant effect on being at or below lower hemoglobin level (severely anemic). Furthermore, age of child (OR=0.92, 95% CI: 0.86-1.00) was a significant predictor, indicating that odds of severe anemia decreases if the child is 48 months or older.

[1] Department of Statistics, St. Anthony's College, Shillong, Meghalaya-793001, India
Email: sanku_dey2k2003@yahoo.co.in
[2] Applied Statistics and Research Methods, University of Northern Colorado, Greeley, Colorado 80639,
E-mail: enayetur.raheem@unco.edu

# INTRODUCTION

There is a saying "A man never stands as tall as when he kneels to help a child". A child is the future of a nation, and if there is no effort made to the wellbeing of the nation's future, then nation building will suffer adversely with major consequences for social and economic development. In this context, we observe in the well documented literature that anemia is the most common nutritional problem both in developed and developing countries, while the prevalence of anemia among pre-school children is 47.4% [1]. About one third of the global population (over two billion) is anemic [2]. According to WHO estimates, India has the highest prevalence of anemia among the South Asian Countries in all age groups. The Third National Family Health Survey [3] of India revealed that 70% children are anemic in the age group of 6–59 months, including 3% severely anemic, 40% moderately anemic, and 26% mildly anemic [4]. As per five major surveys (National Family Health Survey (NFHS 2 & 3), District Level Household Survey 2 (DLHS), Indian Council of Medical Research (ICMR) Micronutrient Survey [5] and Micronutrient Survey [6] over 70 per cent of preschool children are anemic in the country.

Iron deficiency (ID) is listed as one of the "Top Ten Risk Factors contributing to Death" [7]. Iron deficiency anemia (IDA) is prevalent in South Asia, predominantly India, Bangladesh and Pakistan. However, the prevalence of IDA has declined substantially in neighboring countries such as Bangladesh and Pakistan [8], whereas it has plummeted from 20% to 8% within a decade in China [9]. Studies carried out in Egypt [10], India [11], Thailand [12], and the United States [13] revealed that iron and folate deficiency anemia reduces the learning capacity, attentiveness and intelligence of children aged below five years and primary school children. Several studies have been carried out on anemia in India since 1980's. However, very few studies focused on child anemia at the national and regional levels [1,14,15]. Nutritional problem is very common among children below five years in all the states of India, particularly, in those states whose performances are very poor in respect of demographic and socio-economic indicators. There are several factors responsible for IDA, for example, low dietary intake, inadequate iron (less than 20 mg/day) and folic acid intake (less than 70 mg/day), chronic blood loss due to infection such as malaria and hookworm infestations, among others. [5,6]. Therefore, it is imperative to find out the factors that contributed to anemia and to examine the contribution of existing programmes in combating child anemia especially in the less developed areas like northeastern states of India. In view of the above, our objective is to determine the prevalence of anemia among children (0 - 5 years of age) from the northeastern States of India. To comprehend the prevalence of anemia and to identify the significant predictors, socio-economic differentials have been taken into consideration in our study.

## MATERIALS AND METHODS

The present paper uses the dataset from the 2005–2006 National Family Health Survey (released in 2008) for the northeastern states of India. Respondents were selected through a systematic multistage stratified sample survey conducted in all 29 states of India [3]. In each state, populations were stratified by urban and rural area of residence, and the sample size at the state level was proportional to the size of the state's urban and rural population. The study population constitutes a nationally representative cross-sectional sample of singleton children aged 0 to 59 months and born after January 2000 or January 2001 (n=50,750) to mothers aged 15 to 49 years from all 29 states of India. Information on children was obtained by a face-to-face interview with mothers, with a response rate of 94.5%. The survey also provided district-level information on the prevalence of under-nutrition [weight-for-age, using the standard deviation (SD) classification] among children in the age-group of 0-59 month(s); prevalence of anemia (Hb estimation by HemoCue system) for the children aged 6-59 months, adolescent girls aged 10-19 years, pregnant women; household availability of iodized salt; and the coverage of vitamin A programme, with appropriate dosage. To meet the objectives of the study, we have produced a dataset that pertains to the North Eastern States of India. Our study comprises of 10,136 children within 0-5 years of age.

**Variables**

The variables of the study are briefly described in the following.

*Outcome/response variable:* We create the response variable using the variables that measures hemoglobin level in blood. We split the variable into four categories as per the specification of WHO to define level of anemia. In particular, we define a new variable called `hglevel` to indicate level of hemoglobin in blood (g/dL). We set hglevel=1 if hemoglobin level is below 7 g/dL and call is severely low level (indicating severe anemia), hglevel=2 if hemoglobin level is 7-9.9 g/dL (indicating moderate anemia), hglevel=3 if hemoglobin level is 10-10.9 g/dL (indicating mild anemia), and hglevel=4 if hemoglobin level is above 11 g/dL indicating no anemia.

*Independent variables/predictors:* The study includes a set of independent variables to understand the extent and differentials in the level of anemia among children of age group 0-5 years. Table 1 represents an overview of the predictors used in the model.

**Objective of the study / research questions**

Our objective is to study the relationship between some household level predictors and some maternal variables on the probability of being at or below a hemoglobin level. Based on our classification of anemic children, we would then calculate odds ratio and the predictive probability of being at or below a hemoglobin level based on a polytomous logistic regression model.

In particular, we like to answer the following research questions:

1. What is the likelihood of being at or below each level of hemoglobin in the blood (g/dL) for children at a typical household?
2. Does the likelihood of being at or below each level of hemoglobin (i.e., anemia level) vary across state of residence?
3. What are the relationships between the household and maternal variables, and the likelihood of a child being at or below given hemoglobin level?

## STATISTICAL ANALYSIS

In the present study we employ bivariate as well as multivariate techniques to identify the factors that are associated with anemia in children of 0-5 years of age. In the bivariate analysis, cross tabulation was made between the potential risk factors and the presence of anemia. Chi-square test and p-values were used to test the significance of each of the potential risk factor in bivariate analysis. These risk factors are then tested for their relationship with the outcome in multivariate setup. We use multinomial multilevel logistic regression model to predict the level of anemia as a function of mother's age at marriage, number of children ever born to mother, religion, literacy of mother, household living standard, place of residence, sex of the child, and age of child (in months).

Due to the stratified nature of data in NFHS, the children are naturally nested into mothers, mothers are nested into households, households are into Primary Sampling Units (PSUs) and PSUs into states. Hence keeping in view of the hierarchically clustered nature of the survey data, the paper uses multilevel regression model to estimate parameter for anemia among children to avoid the likely under-estimation of parameters from a single level model [16]. The advantage of multilevel model is that they properly account for the correlation structure of the data that frequently occur in social sciences and in multistage survey sampling.

To properly account for the hierarchical nature of our data, we consider state of residence as the level-2 variable under which the respondents (level-1) are nested. Thus, respondent-level variables are the level-1 variables used in this study to predict the level of anemia in children. Our outcome variable is ordinal polytomous with four levels for anemia. The model for such data uses a multinomial distribution and cumulative logit link function to compute the cumulative odds for each category of the response [17].

For the polytomous response variable, let us denote $\eta_{kij}$ as the log odds of a child in the $i$th household in the $j$th state being at or below $k$th level of hemoglobin in blood. The model for level-1 may be written as

$$\eta_{kij} = \log\left(\frac{P(R_{ij} \leq k)}{1 - P(R_{ij} \leq k)}\right) = \beta_{oj} + \beta_{1j} X_{ij} + \delta_k \qquad (1)$$

Here, $\beta_{0j}$ is the intercept, assumed to be random, represents the average log odds of being at or below severely low hemoglobin level; $P(R_{ij} \leq k)$ represents the probability of responding at or below $k$th level of the outcome variable; $X_{ij}$ represents level-1 predictor, $\beta_{1j}$ is the slope coefficient corresponding to $X_{ij}$ which measures the change in the probability of being at a given hemoglobin level per unit change in the level-1 predictor; and $\delta_k$ represents the difference between the $k$th category and the preceding one. Essentially, for $k = 1$, we have $\delta_1 = 0$.

We consider intercept to be random having the following model

$$\beta_{0j} = \gamma_{00} + u_{oj},$$

where $\gamma_{00}$ is the average log odds of a typical household in the population, and $u_{0j}$ is the random error term with $u_{0j} \sim N(0, \tau_{00})$ where $\tau_{00}$ is the level-1 error variance.

Now substituting the expression for $\beta_{0j}$ in equation (1), we get

$$\eta_{kij} = \log\left(\frac{P(R_{ij} \leq k)}{1 - P(R_{ij} \leq k)}\right) = \gamma_{00} + \beta_{1j} X_{ij} + \delta_k + u_{0j} \quad (2)$$

This model represents a scenario where there is no level-2 predictor (as in our case), and one level-1 predictor. However, in our study, we have more than one level-1 predictor, and so the model is meant to be augmented by those additional covariates as they are considered in the analysis.

Model (2) gives log odds when fitted to data. Alternatively, predicted probability of the event of interest (e.g., being at or below a hemoglobin level) can be calculated using the following formula:

Predicted Probability (PP) = $\frac{e^{\eta_{kij}}}{1 + e^{\eta_{kij}}}$ \quad (3)

*Model building strategies for multilevel logistic regression*

We consider multilevel multinomial logistic regression model to predict the probability of being at or below a hemoglobin level using the available risk factors. Since the outcome variable is ordinal, we consider cumulative logit link function. The steps of the model building process are outlined in Table 4. We consider random intercept model where the model is fitted with only the intercept (Model 1). Then in Model 2 we add maternal characteristics such as number of children ever born (2 or less, 3-4, and 5 or more), and age at marriage (below 18, 18-26 and above 26 years of age). In Model 3 we add child's age in months (<48 months, 48 or more months). To get Model 4, we added religion and literacy of mother to Model 3. Finally, we augment Model 4 by adding sex of child, household living standard (low, medium, high), and place of residence (urban vs. rural). We call it Model 5.

# RESULTS

**Bivariate analysis**

Of the 10,136 children who were included in this study 51.6% were male and 48.4% were female children. 52.5% children were anemic. Out of these children, 53% male children and 52% female children were anemic. Tables 2 and 3 present the cross tabulation of prevalence of anemia (frequency and percentages) by level of anemia. Prevalence of anemia with regard to state of residence shows that Tripura has the highest percentage of anemia cases. Of the 708 children in Tripura, only 25% are non-anemic and the remaining 75% have some form of anemia (19% severe, 50% moderate, and 6% mild anemia). Sikkim falls next to Tripura with about 71% anemia cases (22% severe, 43% moderate, and 6% mild). Nagaland and Meghalaya each has 48% anemia cases while Arunachal Pradesh has 45% followed by Manipur which has the lowest percentage of anemia cases (43%). Among the severely anemic children, 27% are in Assam, 26% in Arunachal Pradesh followed by 20% in Manipur. Assam has the most moderately anemic children (35%) followed by Arunachal Pradesh (21%) and Manipur (15%). Among the non-anemic children, 33% are in Arunachal Pradesh followed by 24% in Manipur.

Chi-square test for association shows that the place of residence is associated with the level of anemia ($\chi^2 = 622.84, p < .0001$). In other words, prevalence of anemia among children varies significantly among the states.

Anemia is slightly more prevalent among males (53%) than females (52%). Among the severely anemic children, 85% are in rural areas compared to 15% in urban areas. Similarly, 79% of moderately anemic children, and 75% of mildly anemic children live in rural areas. Muslim children have the highest prevalence rate of any form of anemia (62%) followed by 56% among Hindus. Christians have the lowest prevalence of anemia (48%). However, among the severely anemic child, 44% are Hindu, 7% Muslim, 29% are Christian.

Of the severely anemic cases, 62% are from household with low living standard, 27% are from medium, and 10% are from high living standard households. Interestingly, prevalence of any form of anemia is low among the mothers who cannot read or write. However, the association is not statistically significant ($\chi^2 = 6.43, p = .09$). Among the mothers of severely anemic children, 46% have two or less children, 36% have three to four children, and 18% have more than five children. Majority of the mothers of severely anemic children (63%) were married between the ages 18 and 26 years, 32% were married below 18 years while 5% were married when they were above 26 years of age.

The bivariate analysis suggests that there is statistically significant association between the level of anemia (severe, moderate, mild, and none) and religion of the respondents ($p < .0001$), and place of residence ($p < .0001$), and between level of anemia and age of mother at marriage ($p = .0425$). There is no significant association found between anemia level and household living standard, sex of child, literacy of mother, total children ever born to mother, and age of child (months). Owing to this, we investigate the relationship between anemia level and the variables cited above using multilevel polytomous logistic regression.

**Multivariate Analysis**

In multivariate analysis, the fitted models along with the estimated effects and their standard errors are presented in Table 5. We used Laplace estimation method so that we can compare the models based on negative 2 log-likelihood (-2LL) or the deviance test. The last row of Table 5 shows the p-values for the deviance test based on Chi-square statistic.

Notice that Model 1 is nested under Model 2, which in essence nested under Model 3, and this in turn is nested under Model 4, and again Model 4 nested under Model 5. This allows us to compare these models based on -2LL. We found Model 3 to be better than Model 1 as the drop in -2LL is statistically significant (p < .0001), and Model 3 to be better than Model 2 ($p = .0452$) as well. Models 4 and 5 include more variables and thus reduce the -2LL values further. However, the gain in reduction of -2LL is not statistically significant enough to consider them for our analysis. We, therefore, resort to Model 3 and use this model to answer our research questions.

In regard to our first research question about the likelihood of being at or below given hemoglobin level (indicative of anemia level) at a given household, we consider the fixed effects estimates. The estimated effects are actually the log-odds, which we will use to calculate the predictive probability (PP) of a child being at or below a hemoglobin level. For example, the probability of being at or below severely low hemoglobin level (i.e., severely anemic) for a given child with average background characteristics is P (being at or below severely anemic level) $= \frac{e^{\eta_{ij}}}{1+e^{\eta_{ij}}} = \frac{e^{-1.95}}{1+e^{-1.95}} = \frac{.1423}{1+.1423} = 0.1247$

Similarly, the probability of being at or below moderately anemic level is 0.4825, at or below mildly anemic is 0.5523. These are cumulative probabilities of being at or below a given level of anemia. In order to obtain the exact probability of being at a given level, we have to subtract the adjacent probabilities. For example, the predicted probability of being at or below severely anemic level is 0.1247, at the moderately anemic level is (0.4825-0.1247=) 0.3578, at the mild anemic level is (0.5523-0.4825=) 0.0698. Also, probability of being non-anemic is 1-0.5523 = 0.4477.

In the second research question we wanted to know if the likelihood of being at or below each level of hemoglobin (i.e., anemia level) varies across state of residence. To answer this question, we estimated the covariance parameter for the variable 'state'. The estimated intercept for 'state' is $0.2015$ $(z = 1.94, p = .0263)$. Thus there exists statistically significant variation across states for the likelihood of being at or below a hemoglobin level. Additionally, the intra-class correlation coefficient (ICC) was computed using the covariance parameter estimate for the state (0.2015). The ICC indicates how much of the total variation in the probability of being at or below a hemoglobin level is due to the variation among the states. Following the procedure of Snijders & Bosker [18] we calculate ICC as

$$ICC = \frac{\tau_{00}}{\tau_{00} + 3.29} = \frac{0.2015}{0.2015 + 3.29} = 0.0577$$

The ICC=0.0577 indicates that approximately 6% of the total variation in the response is accounted for by the states. Thus, the remaining 94% variability is due to the variation within the respondents / households and other unknown factors.

Finally, to answer our last research question, we use the best fit model (Model 3) as discussed earlier. Model 3 contains three significant covariates (age at marriage, number of children ever born, and age of child in months). We find significant effect of age at marriage $(\hat{\beta} = 0.12, p = .04)$, number of children ever born $(\hat{\beta} = 0.09, p = .03)$, and age of child $(\hat{\beta} = -0.08, p = .04)$. The estimates for mother's age at marriage and the number children ever born have positive effect on the log odds, whereas age of child has a negative effect. This implies if the age at marriage increases by one unit, the corresponding change in the log odds is 0.12. Similarly, a unit increase in the number of children ever born to mother increases the log odds by 0.09, and getting 48 months or older reduces the log odds by 0.08 units. In other words, as age at marriage and number of children born to women increases the probability of being at the severely anemic level increases. On the other hand increase in age of child decreases the corresponding probability. Since we have four levels for the outcome variables, we have three intercepts as shown in Table 5. The tests for these fixed intercepts are given in Table 6.

We also calculated odds ratio for these predictors. For age at marriage (OR=1.13, 95% CI: 1.05, 1.21), one unit increase in age at marriage results in 1.13 times increase of having lower hemoglobin level (i.e., becoming more anemic). For number of children even born (OR=1.09, 95% CI: 1.03, 1.15), one unit increase in the number of children born to women inflates the odds of having low hemoglobin level (indicating severely anemic) by 1.09. On the other hand, age of child (OR=0.92, 95% CI: 0.86-1.00) indicates odds of being at a lower hemoglobin level decreases if child is 48 months or older.

The predicted probability of being at a given anemia level is shown in Tables 7, 8, and 9 for each of the risk factors.

## DISCUSSION

Our analysis suggests that childhood anemia is highly prevalent in the north eastern states of India. The overall prevalence of anemia in children in the north eastern states is 53%. There are important and significant relationships between anemia and some of the selected potential risk factors. Bivariate analysis suggests that there is no significant association between anemia level and household living standard, sex of child, literacy of mother, total children ever born to mother, and age of child (months) whereas, results from multivariate analysis show that age of the child, mother's age at marriage and number of children ever born have significant effect on child anemia.

In contrary to our findings, several other studies [19,20] indicated that male children were at greater risk of anemia than female children. Also children living in rural areas were at greater risk of anemia as compared to their urban counterparts. Results also show that children belonging to households with low and medium standard of living index are more susceptible to have anemia compared to their counterparts which corroborates the findings in Brazil and other countries [21,22].

Interestingly, the prevalence of any form of anemia is low among the illiterate mothers. Results also indicate that the highest prevalence rate of any form of anemia was amongst the Muslim children (62%) followed by Hindus (56%) and Christians (48%). The reason might be that the children from Muslim and Hindus are from lowest quintiles and illiterate mothers. The fertility of the mother impacted anemia in children i.e., greater the number of children of the mother, higher is the risk of anemia. Mother's age at marriage had also a considerable effect on anemia in their children. Children of women who got married between 18 and 26 years were at greater risk of anemia [20,23].

The results from this study indicate that ignoring the hierarchical nature of the data could result in over statement of the significance of some of the variables included in the model. Most importantly, the standard errors would be biased downward. Anemia is a widespread health problem in India, especially, less developed regions like north eastern states of India. The study findings have some important and relevant policy messages. High prevalence of mild and moderate anemia demands multiple interventions and strategies to tackle the burden of anemia. Special policies must be formulated and executed for prevention of anemia.

The present study demonstrates the endemicity of childhood anemia in northeastern states of India which may be due biological, social and cultural factors. However, in this study mother's age at marriage and the number of children ever born are the factors significantly associated with anemia in children. Results also suggest that age of the child is also a significant predictor, as the age of the child gets 48 months or more, likelihood of severe anemia decreases. This study throws light on the fact that anemia intervention needs to focus more at state level followed by individual level. The limitations of the study are that indicators like, serum ferritin, Dietary intake, worm infestations, malaria, and infectious diseases [24] have not been considered.

Table 1: Risk factors of anemia and their categories.

| Variable name | Variable type | Coding (if categorical) |
|---|---|---|
| State of residence | Categorical | 8 states |
| Place of residence | Categorical | 1=Rural, 2=Urban |
| Religion | Categorical | 1=Hindu, 2=Muslim, 3=Christian, 4=Others |
| Household living standard | Categorical | 1=Low, 2=Medium, 3=High |
| Sex of child | Categorical | 1=Male, 2=Female |
| Literacy of Mother | Categorical | 1=Can read and write, 2=Cannot read and write |
| Total children ever born | Categorical | 1=2 or less, 2=3-4 children, 3=5 or more |
| Age at marriage (mother) | Categorical | 1=18 or less, 2=18-26, 3=26 or more |
| Age of child (months) | Categorical | 1= below 48, 2= 48 or more |

Table 2: Cross-table of prevalence of anemia by state of residence.

| State | Anemia Level | | | | |
|---|---|---|---|---|---|
| | Severely Anemic | Moderately Anemic | Mildly Anemic | Non Anemic | Total |
| **Sikkim** | 159 | 301 | 42 | 206 | 708 |
| | 22.46% | 42.51% | 5.93% | 29.10% | 100% |
| | 12.99% | 8.82% | 6.12% | 4.28% | - |
| **Arunachal Pradesh** | 317 | 724 | 261 | 1611 | 2913 |
| | 10.88% | 24.85% | 8.96% | 55.30% | 100% |
| | 25.90% | 21.21% | 38.05% | 33.48% | - |
| **Nagaland** | 33 | 180 | 35 | 277 | 525 |
| | 6.29% | 34.29% | 6.67% | 52.76% | 100% |
| | 2.70% | 5.27% | 5.10% | 5.76% | - |
| **Manipur** | 241 | 526 | 118 | 1198 | 2083 |
| | 11.57% | 25.25% | 5.66% | 57.51% | 100% |
| | 19.69% | 15.41% | 17.20% | 24.90% | - |
| **Mizoram** | 51 | 258 | 57 | 296 | 662 |
| | 7.70% | 38.97% | 8.61% | 44.71% | 100% |
| | 4.17% | 7.56% | 8.31% | 6.15% | - |
| **Tripura** | 67 | 181 | 20 | 93 | 361 |
| | 18.56% | 50.14% | 5.54% | 25.76% | 100% |
| | 5.47% | 5.30% | 2.92% | 1.93% | - |
| **Meghalaya** | 22 | 55 | 17 | 105 | 199 |
| | 11.06% | 27.64% | 8.54% | 52.76% | 100% |
| | 1.80% | 1.61% | 2.48% | 2.18% | - |
| **Assam** | 334 | 1189 | 136 | 1026 | 2685 |
| | 12.44% | 44.28% | 5.07% | 38.21% | 100% |
| | 27.29% | 34.83% | 19.83% | 21.32% | - |
| **Total** | 1224 | 3414 | 686 | 4812 | 10,136 |
| | - | - | - | - | 100% |
| | 100% | 100% | 100% | 100% | 100% |

Table 3: Frequency distribution of anemia level by the risk factors of anemia. The first column shows chi-square test for independence; the cells represent the frequency (first row), row percentage (second row), and column percentage (third row).

| Variables (Chi-square and p-value) | | Severely Anemic | Moderately Anemic | Mildly Anemic | Non Anemic | Total |
|---|---|---|---|---|---|---|
| **Place of residence** Chi-square = 27.16 $p < .0001$ | Rural | 1035 12.82% 84.56% | 2705 33.50% 79.23% | 516 6.39% 75.22% | 3818 47.29% 79.34% | 8074 100% - |
| | Urban | 189 9.17% 15.44% | 709 34.38% 20.77% | 170 8.24% 24.78% | 994 48.21% 20.66% | 2062 100% - |
| **Religion** Chi-square=166.35 $p < .0001$ | Hindu | 537 14.24% 43.87% | 1330 35.26% 38.96% | 238 6.31% 34.69% | 1667 44.19% 34.64% | 3772 100 - |
| | Muslim | 83 8.99% 6.78% | 442 47.89% 12.95% | 48 5.20% 7.00% | 350 37.92% 7.27% | 923 100% - |
| | Christian | 352 10.14% 28.76% | 1050 30.25% 30.76% | 245 7.06% 35.71% | 1824 52.55% 37.91% | 3471 100% - |
| | Others | 252 12.79% 20.59% | 592 30.05% 17.34% | 155 7.87% 22.59% | 971 49.29% 20.18% | 1970 100% - |
| **Household living standard** Chi-square=8.33 $p = 0.2146$ | Low | 764 12.42% 62.42% | 2031 33.02% 59.49% | 400 6.50% 58.31% | 2956 48.06% 61.43% | 6151 100% - |
| | Medium | 336 11.81% 27.45% | 999 35.10% 29.26% | 199 6.99% 29.01% | 1312 46.10% 27.27% | 2846 100% - |
| | High | 124 10.89% 10.13% | 384 33.71% 11.25% | 87 7.64% 12.68% | 544 47.76% 11.31% | 1139 100% - |
| **Sex of child** Chi-square=2.08 $p = 0.5548$ | Male | 636 12.16% 51.96% | 1772 33.89% 51.90% | 368 7.04% 53.64% | 2453 46.91% 50.98% | 5229 100% - |
| | Female | 588 11.98% 48.04% | 1642 33.46% 48.10% | 318 6.48% 46.36% | 2359 48.07% 49.02% | 4907 100% - |
| **Literacy of Mother** Chi-square=6.43 $P = .0923$ | Can read & write | 692 11.78% 58.20% | 2024 34.47% 61.69% | 385 6.56% 58.25% | 2771 47.19% 59.75% | 5872 100% - |
| | Can't read & write | 497 12.75% 41.80% | 1257 32.26% 38.31% | 276 7.08% 41.75% | 1867 47.91% 40.25% | 3897 100% - |

Table 3 (Continued): Frequency distribution of anemia level by the risk factors of anemia. The first column shows chi-square test for independence; the cells represent the frequency (first row), row percentage (second row), and column percentage (third row).

| Variables (Chi-square and p-value) | | Severely Anemic | Moderately Anemic | Mildly Anemic | Non Anemic | Total |
|---|---|---|---|---|---|---|
| **Total children ever born to mother** Chi-square=9.72 $p = .1366$ | Up to Two Children) | 548 12.80% 46.09% | 1444 33.72% 44.01% | 276 6.45% 41.75% | 2014 47.03% 43.42% | 4282 100% - |
| | Three or Four Children | 423 11.39% 35.58% | 1214 32.70% 37.00% | 271 7.30% 41.00% | 1805 48.61% 38.92% | 3713 100% - |
| | Five or Above Children | 218 12.29% 18.33% | 623 35.12% 18.99% | 114 6.43% 17.25% | 819 46.17% 17.66% | 1774 100% - |
| **Age at marriage** Chi-square=13.03 $p = .0425$ | Below 18 Years | 384 12.23% 32.30% | 1016 32.37% 30.97% | 183 5.83% 27.69% | 1556 49.57% 33.55% | 3139 100% - |
| | 18 To 26 Years | 748 12.20% 62.91% | 2090 34.10% 63.70% | 445 7.26% 67.32% | 2846 46.43% 61.36% | 6129 100% - |
| | Above 26 Years | 57 11.38% 4.79% | 175 34.93% 5.33% | 33 6.59% 4.99% | 236 47.11% 5.09% | 501 100% - |
| **Age of child in months** Chi-square=5.07 $p = .1669$ | >48 months | 499 11.54% 40.77% | 1477 34.17% 43.26% | 313 7.24% 45.63% | 2034 47.05% 42.27% | 4323 100% - |
| | <48 months | 725 12.47% 59.23% | 1937 33.32% 56.74% | 373 6.42% 54.37% | 2778 47.79% 57.73% | 5813 100% - |

Table 4: Model building strategies

| Model 1 | Model 2 | Model 3 | Model 4 | Model 5 |
|---|---|---|---|---|
| No predictors but only random effects for the 'state' | Model 1 + level-1 fixed effects that are related to maternal characteristics such as number of children ever born (2 or less, 3-4 and 5 or more), and age at marriage (below 18, 18-26, and above 26 years). | Model 2+age of child (months) | Model 3 + level-1 fixed effects religion and literacy of mother | Model 4 + sex of child (male/female), household living standard (low, medium, high), and place of residence (urban/rural) |
| Results used to determine the percentage variation in anemia level explained by the level-2 units (state of residence) | Results indicate the relationship between level-1 (household level) predictors and the anemia level | Results indicate if addition of child's age improves the fit. | Results indicate whether addition of demographic variables improves the model fit | Results indicate whether addition of SES variables improves the model fit |

Table 5: Estimates for two-level multinomial logistic regression models for predicting anemia level (N=10,136)

|  | Model 1 | Model 2 | Model 3 | Model 4 | Model 5 |
|---|---|---|---|---|---|
| *Fixed Effects* | | | | | |
| Intercept 1 (Severely anemic) | -1.95* (0.16) | -2.29* (0.19) | -2.35* (0.19) | -2.42* (0.20) | -2.23* (0.23) |
| Intercept 2 (Moderately anemic) | -0.07 (0.16) | -0.42 (0.19) | -0.48* (0.19) | -0.55 (0.20) | -0.37 (0.23) |
| Intercept 3 (Mildly anemic) | 0.21 (0.16) | -0.14 (0.19) | -0.19 (0.19) | -0.27 (0.20) | -0.08 (0.23) |
| Age at marriage | | 0.12* (0.04) | 0.12* (0.04) | 0.12* (0.04) | 0.12* (0.04) |
| Total children ever born | | 0.08* (0.03) | 0.09* (0.03) | 0.08* (0.03) | 0.08* (0.03) |
| Age of child (<48 months) | | | -0.08* (0.04) | 0.08* (0.04) | 0.08* (0.04) |
| Religion | | | | 0.03 (0.02) | 0.02 (0.02) |
| Literacy of mother | | | | 0.02 (0.04) | 0.003 (0.04) |
| Household living standard | | | | | -0.02 (0.03) |
| Place of residence | | | | | -0.06 (0.05) |
| Sex of child | | | | | -0.03 (0.04) |
| *Error Variance* | | | | | |
| Intercept | 0.2015* (0.10) | | | | |
| *Model Fit* | | | | | |
| -2LL | 23,063.22 | 22,228.13** (p<.0001) | 22,224.12** (p=.0452) | 22,222.03 (p=.3517) | 22,218.83(p=.3618) |

Entries in the table are estimated effects while the standard errors are reported in the parenthesis.
* indicates $p < .05$; ** significant LR test; ICC=0.0577. PROC GLIMMIX in SAS 9.4 with Laplace estimation method was used.

Table 6: Tests for the intercepts in Model 1.

| Effect | hglevel | Estimate | Standard Error | DF | t Value | Pr > |t| | Alpha | 95% CI Lower | 95% CI Upper |
|---|---|---|---|---|---|---|---|---|---|
| **Intercept** | Severely Anemic | -1.9478 | 0.1628 | 7 | -11.96 | <.0001 | 0.05 | -2.3328 | -1.5629 |
| **Intercept** | Moderately Anemic | -0.0735 | 0.1612 | 7 | -0.46 | 0.6621 | 0.05 | -0.4548 | 0.3077 |
| **Intercept** | Mildly Anemic | 0.2102 | 0.1612 | 7 | 1.30 | 0.2335 | 0.05 | -0.1710 | 0.5915 |

Table 7: Predicted probability of being at a given anemic level by age at marriage

| | Age at marriage | | |
|---|---|---|---|
| Anemia level | <18 | 18-26 | >26 |
| Severely anemic | 0.0871 | 0.0971 | 0.1081 |
| Moderately anemic | 0.2952 | 0.3138 | 0.3322 |
| Mild anemic | 0.0703 | 0.0716 | 0.0722 |
| Non-anemic | 0.5474 | 0.5175 | 0.4875 |

Table 8: Predicted probability of being at a given anemic Level by number of children ever born to mother.

| | Number of Children Ever Born to Mother | | |
|---|---|---|---|
| Anemia level | 2 or less | 3-4 | 5 or more |
| Severely anemic | 0.0871 | 0.0945 | 0.1025 |
| Moderately anemic | 0.2952 | 0.3092 | 0.3231 |
| Mild anemic | 0.0703 | 0.0713 | 0.0719 |
| Non-anemic | 0.5474 | 0.5250 | 0.5025 |

Table 9: Predicted Probability of being at a given anemic level by age of child (months).

| | Age of Child (months) | |
|---|---|---|
| Anemia level | <48 | 48 or older |
| Severely anemic | 0.0871 | 0.0809 |
| Moderately anemic | 0.2952 | 0.2826 |
| Mild anemic | 0.0703 | 0.0694 |
| Non-anemic | 0.5474 | 0.5671 |